# MOONS: The New Multi-Object Spectrograph for the VLT

Michele Cirasuolo[1,a]
Alasdair Fairley[2]
Phil Rees[2]
Oscar A. Gonzalez[2]
William Taylor[2]
Roberto Maiolino[3]
Jose Afonso[4]
Chris Evans[2]
Hector Flores[6]
Simon Lilly[5]
Ernesto Oliva[7]
Stephane Paltani[8]
Leonardo Vanzi[9]
Manuel Abreu[4]
Matteo Accardo[1]
Nathan Adams[15]
Domingo Álvarez Méndez[1]
Jean-Philippe Amans[6]
Stergios Amarantidis[4]
Hakim Atek[6]
David Atkinson[2]
Manda Banerji[12]
Joe Barrett[2]
Felipe Barrientos[9]
Franz Bauer[9]
Steven Beard[2]
Clementine Béchet[9]
Andrea Belfiore[11]
Michele Bellazzini[13]
Christophe Benoist[32]
Philip Best[10]
Katia Biazzo[14]
Martin Black[2]
David Boettger[9]
Piercarlo Bonifacio[6]
Rebecca Bowler[15]
Angela Bragaglia[13]
Saskia Brierley[2]
Jarle Brinchmann[16]
Martin Brinkmann[1]
Veronique Buat[17]
Fernando Buitrago[4]
Denis Burgarella[17]
Ben Burningham[18]
David Buscher[3]
Alexandre Cabral[4]
Elisabetta Caffau[6]
Leandro Cardoso[16]
Adam Carnall[10]
Marcella Carollo[5]
Roberto Castillo[1]
Gianluca Castignani[19]
Marco Catelan[9]
Claudia Cicone[20]
Andrea Cimatti[21]
Maria-Rosa L. Cioni[22]
Gisella Clementini[13]
William Cochrane[2]
João Coelho[4]
Miriam Colling[23]
Thierry Contini[24]
Rodrigo Contreras[9]
Ralf Conzelmann[1]
Giovanni Cresci[7]
Mark Cropper[25]
Olga Cucciati[13]
Fergus Cullen[10]
Claudio Cumani[1]
Mirko Curti[3]
Antonio Da Silva[4]
Emanuele Daddi[26]
Emanuele Dalessandro[13]
Francesco Dalessio[14]
Louise Dauvin[9]
George Davidson[2]
Patrick de Laverny[32]
Françoise Delplancke-Ströbele[1]
Gabriella De Lucia[27]
Ciro Del Vecchio[7]
Miroslava Dessauges-Zavadsky[8]
Paola Di Matteo[6]
Herve Dole[28]
Holger Drass[9]
Jim Dunlop[10]
Rolando Dünner[9]
Steve Eales[29]
Richard Ellis[30]
Bruno Enriques[5]
Giles Fasola[6]
Annette Ferguson[10]
Debora Ferruzzi[7]
Martin Fisher[3]
Mauricio Flores[9]
Adriano Fontana[14]
Vincenzo Forchi[1]
Patrick Francois[6]
Paolo Franzetti[11]
Adriana Gargiulo[11]
Bianca Garilli[11]
Julien Gaudemard[6]
Mark Gieles[31]
Gerry Gilmore[3]
Michele Ginolfi[8]
Jean Michel Gomes[16]
Isabelle Guinouard[6]
Pablo Gutierrez[1]
Régis Haigron[6]
François Hammer[6]
Peter Hammersley[1]
Chris Haniff[3]
Chris Harrison[1]
Misha Haywood[6]
Vanessa Hill[32]
Norbert Hubin[1]
Andrew Humphrey[16]
Rodrigo Ibata[33]
Leopoldo Infante[9]
Derek Ives[1]
Rob Ivison[1]
Olaf Iwert[1]
Pascale Jablonka[19]
Gerd Jakob[1]
Matt Jarvis[15]
David King[3]
Jean-Paul Kneib[19]
Philippe Laporte[6]
Andy Lawrence[10]
David Lee[2]
Gianluca Li Causi[35]
Silvio Lorenzoni[4]
Sara Lucatello[34]
Yerco Luco[9]
Alastair Macleod[2]
Manuela Magliocchetti[35]
Laura Magrini[7]
Vincenzo Mainieri[1]
Charles Maire[8]
Filippo Mannucci[7]
Nicolas Martin[33]
Israel Matute[4]
Sophie Maurogordato[32]
Sean McGee[43]
Derek Mcleod[10]
Ross McLure[10]
Richard McMahon[3]
Basile-Thierry Melse[6]
Hugo Messias[4]
Alessio Mucciarelli[21]
Brunella Nisini[14]
Johannes Nix[2]
Peder Norberg[36]
Pascal Oesch[8]
António Oliveira[4]
Livia Origlia[13]
Nelson Padilla[9]
Ralf Palsa[1]
Elena Pancino[7]
Polychronis Papaderos[16]
Ciro Pappalardo[4]
Ian Parry[3]
Luca Pasquini[1]
John Peacock[10]
Fernando Pedichini[14]
Roser Pello[17]
Yingjie Peng[42]
Laura Pentericci[14]
Oliver Pfuhl[1]
Roberto Piazzesi[14]
Dan Popovic[1]
Lucia Pozzetti[13]
Mathieu Puech[6]
Thomas Puzia[9]
Anand Raichoor[19]
Sofia Randich[7]
Alejandra Recio-Blanco[32]
Sandra Reis[4]
Florent Reix[6]

Alvio Renzini[34]
Myriam Rodrigues[6]
Felipe Rojas[9]
Álvaro Rojas-Arriagada[9]
Stefano Rota[11]
Frédéric Royer[6]
Germano Sacco[7]
Ruben Sanchez-Janssen[2]
Nicoletta Sanna[7]
Pedro Santos[4]
Marc Sarzi[44]
Daniel Schaerer[8]
Ricardo Schiavon[37]
Robin Schnell[8]
Mathias Schultheis[32]
Marco Scodeggio[11]
Steve Serjeant[38]
Tzu-Chiang Shen[39]
Charlotte Simmonds[19]
Jonathan Smoker[1]
David Sobral[45]
Michael Sordet[8]
Damien Spérone[19]
Jonathan Strachan[2]
Xiaowei Sun[3]
Mark Swinbank[36]
Graham Tait[2]
Ismael Tereno[4]
Rita Tojeiro[40]
Miguel Torres[9]
Monica Tosi[13]
Andrea Tozzi[7]
Ezequiel Tresiter[9]
Elena Valenti[1]
Álvaro Valenzuela Navarro[9]
Eros Vanzella[13]
Susanna Vergani[6]
Anne Verhamme[19]
Joël Vernet[1]
Cristian Vignali[13]
Jakob Vinther[1]
Lauren Von Dran[41]
Chris Waring[2]
Stephen Watson[2]
Vivienne Wild[40]
Bart Willesme[2]
Brian Woodward[2]
Stijn Wuyts[46]
Yanbin Yang[6]
Gianni Zamorani[13]
Manuela Zoccali[9]
Asa Bluck[3]
James Trussler[3]

[1] ESO
[2] STFC, UK Astronomy Technology Centre, Royal Observatory Edinburgh, UK
[3] Cavendish Laboratory, University of Cambridge, UK
[4] Instituto de Astrofísica e Ciências do Espaço and Departamento de Física, Faculdade de Ciências, Universidade de Lisboa, Portugal
[5] Department of Physics, ETH Zurich, Switzerland
[6] GEPI, Observatoire de Paris, PSL University, CNRS, France
[7] INAF-Osservatorio Astrofisico di Arcetri, Florence, Italy
[8] Department of Astronomy, University of Geneva, Versoix, Switzerland
[9] Pontificia Universidad Católica de Chile, Santiago, Chile
[10] Institute for Astronomy, University of Edinburgh, Royal Observatory, Edinburgh, UK
[11] INAF, IASF-MI, Milano, Italy
[12] Faculty of Engineering and Physical Sciences, University of Southampton, UK
[13] INAF – Astrophysics and Space Science Observatory Bologna, Italy
[14] INAF – Osservatorio Astronomico di Roma, Italy
[15] Department of Physics, University of Oxford, UK
[16] Instituto de Astrofísica e Ciências do Espaço, Universidade do Porto, CAUP, Porto, Portugal
[17] Aix Marseille Univ, CNRS, CNES, LAM, Marseille, France
[18] School of Physics Astronomy and Mathematics, University of Hertfordshire, UK
[19] EPFL, Observatoire de Sauverny, Versoix, Switzerland
[20] Institute of Theoretical Astrophysics, University of Oslo, Norway
[21] University of Bologna, Department of Physics and Astronomy (DIFA), Italy
[22] Leibniz-Institut für Astrophysik Potsdam (AIP), Germany
[23] STFC, Daresbury Laboratory, Sci-Tech Daresbury, UK
[24] Institut de Recherche en Astrophysique et Planétologie, Toulouse, France
[25] UCL Department of Space and Climate Physics, London, UK
[26] CEA, IRFU, DAp, AIM, Université Paris-Saclay, Université Paris Diderot, France
[27] INAF – Osservatorio Astronomico di Trieste, Italy
[28] Institut d'Astrophysique Spatiale, Orsay, Université Paris Sud, France
[29] School of Physics and Astronomy, University of Cardiff, UK
[30] Dept of Physics & Astronomy, University College London, UK
[31] Astrophysics Research Group, Surrey University, UK
[32] Université Côte d'Azur, Observatoire de la Côte d'Azur, CNRS, Laboratoire Lagrange, France
[33] Observatoire Astronomique, Université de Strasbourg, France
[34] INAF – Osservatorio Astronomico di Padova, Italy
[35] INAF – Istituto di Astrofisica e Planetologia Spaziali, Rome, Italy
[36] Department of Physics, Durham University, UK
[37] Liverpool John Moores University, UK
[38] School of Physical Sciences, The Open University, Milton Keynes, UK
[39] BlueShadows Ltda., Santiago, Chile
[40] School of Physics and Astronomy, University of St Andrews, UK
[41] Appleton Laboratory, STFC, Harwell Campus, UK
[42] Kavli Institute for Astronomy and Astrophysics, Peking University, Beijing, China
[43] School of Physics and Astronomy, University of Birmingham, UK
[44] Armagh Observatory & Planetarium, Armagh, Northern Ireland
[45] Department of Physics, University of Lancaster, UK
[46] Department of Physics, University of Bath, UK

**MOONS is the new Multi-Object Optical and Near-infrared Spectrograph currently under construction for the Very Large Telescope (VLT) at ESO. This remarkable instrument combines, for the first time, the collecting power of an 8-m telescope, 1000 fibres with individual robotic positioners, and both low- and high-resolution simultaneous spectral coverage across the 0.64–1.8 μm wavelength range. This facility will provide the astronomical community with a powerful, world-leading instrument able to serve a wide range of Galactic, extragalactic and cosmological studies. Construction is**

now proceeding full steam ahead and this overview article presents some of the science goals and the technical description of the MOONS instrument. More detailed information on the MOONS surveys is provided in the other dedicated articles in this Messenger issue.

## Introduction

Over the last two decades several observational milestones have dramatically changed our knowledge of the Universe. Measurements of the Cosmic Microwave Background, high-redshift supernovae and large-scale structure have revealed that 96% of the density of the Universe consists of currently unexplained Dark Energy and Dark Matter, and less than 4% is in the form of baryons. Yet most of the information we have comes from luminous, baryonic matter. Understanding the nature of the dark components which dominate the global expansion and large-scale structure of the Universe along with the physical processes that affect baryons and shape the formation and evolution of stars and galaxies is amongst the most fundamental unsolved problems in science.

Answering these important questions requires an accurate reconstruction of the assembly history of stars and galaxies over virtually all of cosmic time in order to decode the building blocks of the Universe. The Milky Way offers a unique opportunity to reconstruct the assembly history of a prototypical spiral galaxy by looking at the individual ages, chemical abundances, and orbital motions of its stellar populations. Looking far beyond our Galaxy, it is also essential to trace the evolution of galaxy properties (star formation, chemical enrichment, mass assembly, etc.) over the whole cosmic epoch if we are to investigate the effects of age and environment. Ideally, these studies should be pushed to the highest redshifts — when the Universe was just a few hundred million years old — and young galaxies are key to understanding the physics of the early Universe and cosmic re-ionisation. Addressing these fundamental science goals requires accurate determinations of stellar and galactic physical properties, as well as precise measurements of the spatial and chemical distribution of stars in the Milky Way and the 3D distribution of galaxies at different epochs.

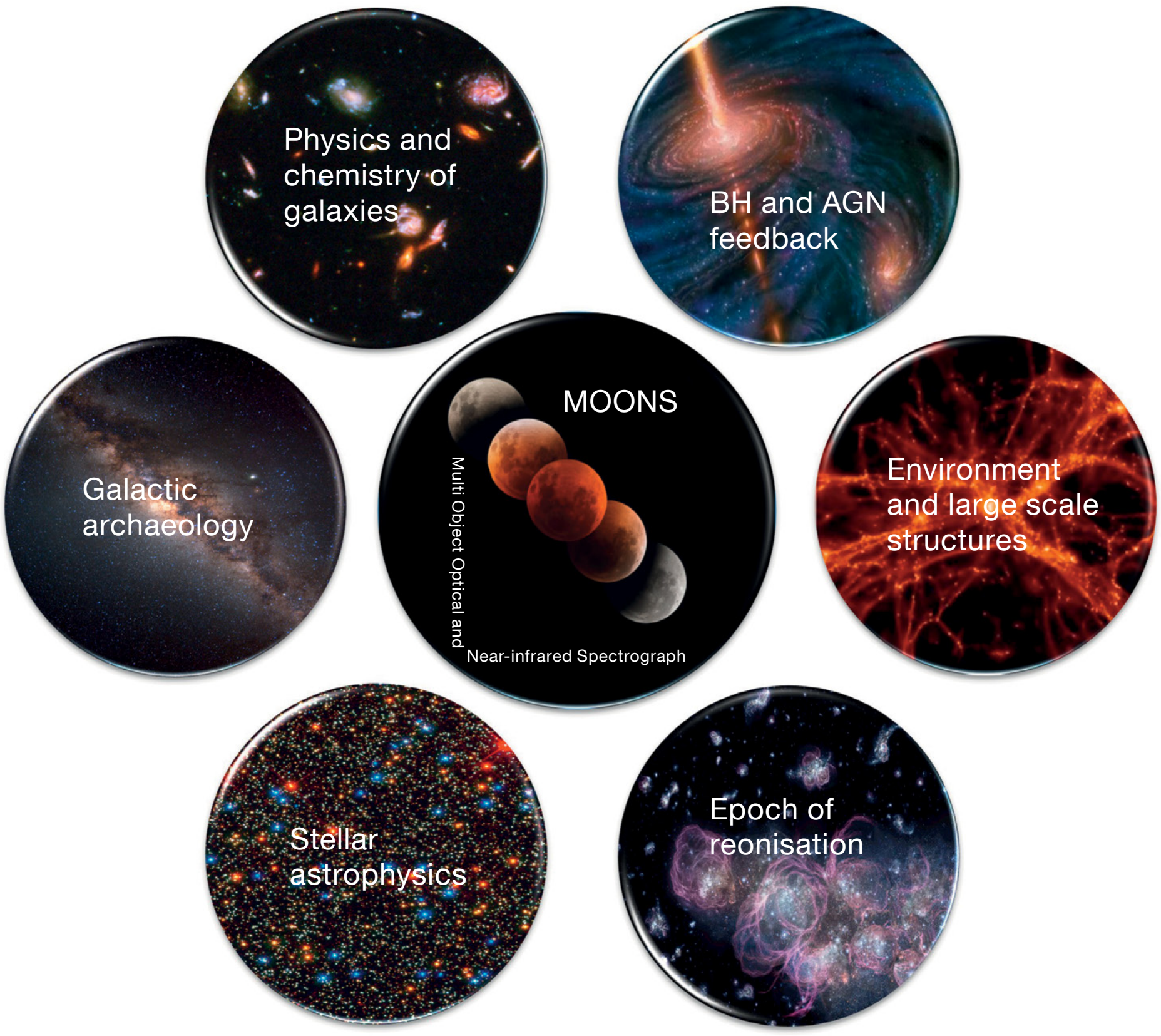

Figure 1. The key science drivers that have shaped the requirements of the MOONS instrument:
- The ability to obtain radial velocities and detailed stellar abundances for millions of stars, especially in the obscured regions of the Galaxy, to reconstruct the chemo-dynamical properties of our Milky Way.
- The capability to observe key spectral diagnostics for millions of distant galaxies up to the epoch of re-ionisation at $z > 7$, to determine the physical processes that shape their evolution and the impact of central supermassive black holes and the environments in which galaxies live, disentangling nature versus nurture effects.

In order to address this fundamental science, the MOONS instrument has been developed focusing on three essential parameters: sensitivity, multiplexing and wavelength coverage (the full list of key instrument parameters is given in Table 1)[1,2]. In respect of sensitivity, the VLT is currently one of the largest infrared and visible telescopes in the world in terms of collecting area, and every element of the MOONS instrument itself has been optimised for high transmission. The multiplex of 1000 is a factor of 20 larger than current spectrographs operating in the near-infrared; this is limited by the budget available, the capability to manufacture very large optics, and space on the Nasmyth platform for the large cryostat (which is already more than 4 m high — see Figure 2). Finally, the broad wavelength coverage of MOONS, from 0.645 µm to 1.8 µm, extending into the near-infrared, is critical for observing heavily dust-obscured regions of our Milky Way as well as for opening a window onto the high-redshift Universe. To meet the aspirations of both the Galactic and extragalactic scientific communities, MOONS offers both low- and high-resolution spectroscopy. In the low-resolution mode ($R$ ~ 4000–7000), the entire 0.645–1.8 µm range is observed simultaneously across the *RI*, *YJ* and *H* atmospheric windows. In the high-resolution mode the *YJ* channel remains unchanged at $R$ ~ 4000, while the two high-resolution dispersers are inserted in the RI and H bands: one with $R$ > 9000 around the Ca triplet region to measure stellar radial velocities, and another with *R* ~ 19 000 in the *H* band for detailed measurements of chemical abundances.

Table 1. MOONS key instrument parameters.

| Parameter | Value |
|---|---|
| Telescope | VLT, 8 m |
| Field of view | 25 arcminutes in diameter |
| Multiplex | 1001 |
| On-sky aperture of each fibre | 1.2 arcseconds |
| Field coverage | > 3 fibres can reach any point in the focal plane |
| # of fibres within a 2-arcminute diameter | 7 |
| Minimum fibre separation | 10 arcseconds |
| Spectral channels | *RI*, *YJ* and *H* bands observed simultaneously |
| Resolution modes | Low and high resolution |
| Low-res simultaneous spectral coverage | 0.64 – 1.8 μm |
| Low-res spectral resolution | $R_{RI}$ = 4100, $R_{YJ}$ = 4300, $R_H$ = 6600 |
| High-res simultaneous spectral coverage | $\lambda_{RI}$ = 0.76 – 0.89 μm, $\lambda_{YJ}$ = 0.93 – 1.35 μm, $\lambda_H$ = 1.52 – 1.64 μm |
| High-res spectral resolution | $R_{RI}$ = 9200, $R_{YJ}$ = 4300, $R_H$ = 19700 |
| Throughput | > 30% in low resolution, > 25% in high resolution |
| Sensitivity (point sources) in 1 hr integration | See Figure 3 for details |
| Continuum high res | S/N > 60 at $H_{AB}$ ~ 17 and $RI_{AB}$ ~ 17.5 |
| Continuum low res | S/N > 5 at mag(AB) ~ 23 rebinning to *R* = 1000 after sky subtraction |
| Emission lines | S/N > 5 for a line flux of > 2 × $10^{-17}$ erg $s^{-1}cm^{-2}$, FWHM = 200 km $s^{-1}$ |
| Calibration methods | Daytime flat fields, attached flats as part of observations, ThAr lamps for wavelengths |
| Observing overheads | Fibre positioning time < 2 mins Attached flats + 2 mins |
| Acquisition star limiting mag | V ~ 21 mag (in 30 sec exposure) |

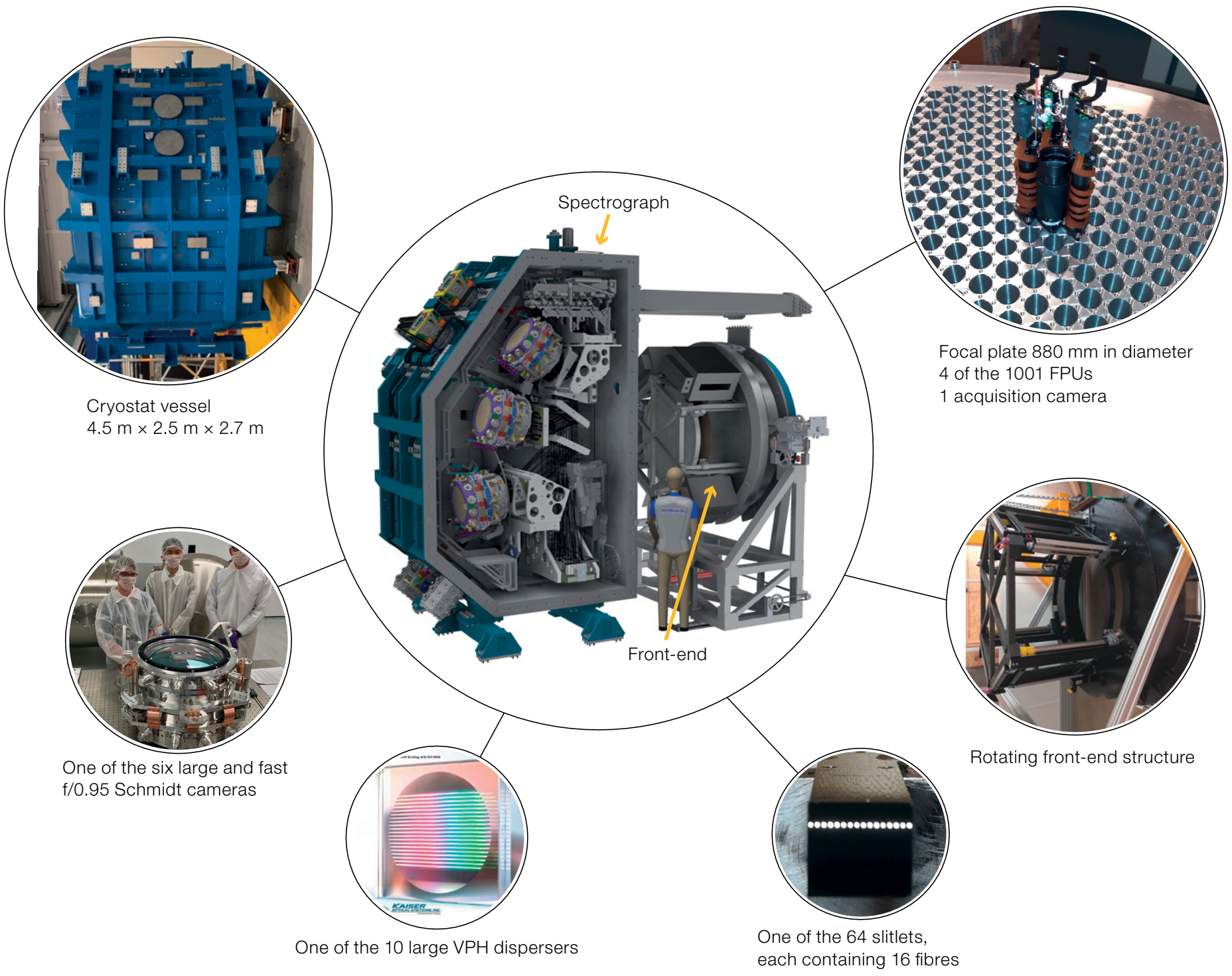

Figure 2. The central figure shows the computer-aided design (CAD) model of the MOONS instrument; the rotating front end and the spectrograph are highlighted. The other images show the real hardware being integrated. From top right and clockwise: the focal plate with four mounted fibre positioning units (FPUs) and one acquisition camera; the structure of the rotating front end that will host the plate with the FPUs, the calibration unit and the metrology system; one of the slitlets with 16 mounted fibres; one of the 10 high efficiency VPH dispersers; one of the six Schmidt cameras; and the cryostat vessel.

## The science drivers and the legacy value

The wealth of science that a highly multiplexed, near-infrared spectrograph like MOONS can generate is undeniably vast and it has been a common aspiration within the ESO community for a long time. MOONS will fill a crucial gap in discovery space which could never be addressed by only optical spectroscopy or low-multiplex near-infrared spectroscopy.

Within the 300 nights of Guaranteed Time Observations (GTO) obtained in return for building the instrument, the MOONS Consortium has developed a coherent set of surveys covering a large fraction of the history of the Universe, from cosmic dawn (13 billion years ago) to the present epoch, across many astrophysical fields (see Figure 1). About 100 GTO nights are devoted to Galactic surveys (see Gonzalez et al., p. 18). The aim is to investigate the nature of the heavily obscured regions of the Galactic bulge (unachievable with optical spectrographs), as well as providing new insights into the chemo-dynamical structure of the thin and thick Galactic discs, and for targeted studies of satellites and streams in the halo.

The other ~ 200 GTO nights will focus on galaxy evolution across cosmic time (see Maiolino et al., p. 24). The goal is to provide a complete picture of the integrated properties of the stellar populations and the ionised interstellar medium (ISM) of galaxies up to high redshift in a SDSS-like survey, including a large number of Lyα emitting galaxies up to *z* ~ 10, and use this to investigate in a systematic way the role that environment and black hole feedback have on the formation and evolution of galaxies with redshift.

The combination of MOONS GTO surveys and open-time surveys will provide an invaluable legacy. Even under the conservative assumption that MOONS is used only for 100 nights a year (i.e., sharing the telescope equally with the other 2 instruments mounted on the UT), it will offer the scientific community ~ 1 000 000 fibre-hours every year. This figure will be even more if in the future one of the UTs is operated in survey mode. On a timescale of 10 years, which is the very minimum lifetime of the instrument, Legacy Surveys with MOONS will provide radial velocities and detailed chemical abundances for tens of millions of stars in our Galaxy and beyond, as well as spectra for millions of galaxies at $0 < z < 10$, providing key spectral diagnostics and environmental information. This will produce a huge and unique dataset of high-quality spectra and the essential deep spectroscopic follow-up of current and future optical and near-infrared imaging surveys or facilities (for example, Gaia, VISTA, UKIDSS, VST, Pan-STARRS, Dark Energy Survey, LSST, Euclid), as well as of objects observed at other wavelengths using, for example, ALMA, Herschel, eRosita, LOFAR, WISE, ASKAP, MeerKAT, etc. Last but not least, MOONS will offer a unique mine from which targets will be selected for detailed follow-up with ESO's Extremely Large Telescope for years to come.

## The MOONS instrument

MOONS is a fibre-fed spectrograph designed to use the full 25-arcminute-diameter field of view (FoV) of one of the Unit Telescopes (UT) of the VLT. The instrument consists of the three major sub-systems shown in Figure 2: the part that is mechanically attached to the telescope and couples the light into the optical fibres (called the rotating front end); the two triple-arm spectrographs — in which the light from the fibres is dispersed and recorded; and the instrument control.

The first element in the optical path of the instrument is the field corrector made of two large lenses of almost 1 m in diameter (and ~ 110 mm thick), which provides a fully corrected field of 25 arcminutes in diameter; this is the largest field possible at the VLT.

The fibres for science observations are deployed on the focal plane created by the field corrector using 1001 miniature fibre positioning units (FPUs), allowing us to configure an entire observation in less than two minutes. Each fibre is connected to its own pick-off unit, which has a footprint of 25 mm fixed on the focal plane and is equipped with two rotating arms[3]. The combination of the two rotations (like the combined motion of elbow and shoulder) allows the fibre to patrol an area with a diameter of 50 mm (~ 1.5 arcminutes on the sky), with an accuracy of better than 20 μm (i.e., less than a third of the diameter of a human hair), which corresponds to 0.05 arcseconds on the sky. The FPUs will be able to achieve this positioning accuracy using state-of-the art stepper motors, but in order to monitor this and make any calibration adjustments there is also an external metrology system capable of precisely measuring the position of each fibre. On the focal plane, embedded within the FPUs, there are 20 acquisition and secondary guiding cameras used to acquire the science field and do a fine alignment of the instrument on the sky. The rotating front end also hosts a novel concept of calibration unit, which uses a projector (like those in cinemas) to illuminate a screen coated in a Lambertian diffuser to guarantee high quality wavelength calibration and flat fielding for all fibres. Indeed, to ensure excellent sky subtraction it is critical that the relative transmission of all the fibres is known very accurately, to better than 1%, and this highly homogeneous illumination is achieved via the calibration unit.

Once the light from stars and galaxies is collected at the front end it is then fed through the fibres to two identical triple-arm spectrographs enclosed in a single cryostat vessel that keeps the optical elements inside at a temperature of −130 degrees C in order to reduce the background in the near-infrared.

In each of the two spectrographs the light from 512 fibres — arranged in 32 slitlets[b], each containing 16 fibres (see Figure 2) — is split by dichroic filters into three wavelength ranges or channels (*RI*, *YJ* and *H*). Each of the two MOONS spectrographs has five highly efficient volume phase holographic (VPH) dispersers, three for the low-resolution mode and two for the high-resolution mode. The two triple-arm spectrographs are mounted back-to-back on the optical bench, which makes it possible to switch between the high- and low-resolution modes in the *RI* band (and similarly for the *H* band) using a single common linear mechanism that passes straight through the optical bench. In each channel, the light dispersed by the VPHs is refocused by using the fastest large cameras ever built for astronomy (to our knowledge). Indeed, these Schmidt-like cameras have a very fast *f*-number of *f*/0.95. Each camera is also very compact and made of just two lenses (glued one inside the other) and one mirror to bring the image into focus on a detector (see Figure 2), and is therefore easy to align. Finally, the light — which has travelled for billions of years in some cases — will be recorded on state-of-the-art detectors. The two infra-red channels (*YJ* and *H*) will exploit the new Hawaii 4RGs 15 μm-pixel detectors and the optical channel (*RI*) will use fully depleted Lawrence Berkeley National Laboratory (LBNL) red-sensitive CCDs.

Since the very beginning of the project, the focus (and the challenge to the engineering team) has been to maximise the quality and the throughput of the instrument or, in other words, the mantra has been "transmit as many photons as physically possible!". For this reason, all the components described above have been optimised in terms of design, material, coating etc. to reach the high sensitivity shown in Figure 3.

## Observation preparation

For complex instruments and particularly for multi-object spectrometers, the usual ESO p2 software used to prepare the observations is complemented by the addition of instrument-specific detailed configuration software. The observation preparation software called MOONLIGHT will perform automatic allocation of fibres to science targets, including optimisation of fibre allocations and accounting for mechanical constraints of the positioners. In order to have high allocation efficiency of the fibres on targets, some overlap between neighbouring patrol fields is needed, with one fibre being able to patrol up to the centre of the neighbouring cell. However, this feature can increase the chances of collisions during positioning. To avoid such collisions, we have developed an algorithm for the path analysis, which calculates in advance the best trajectory and motion of each motor.

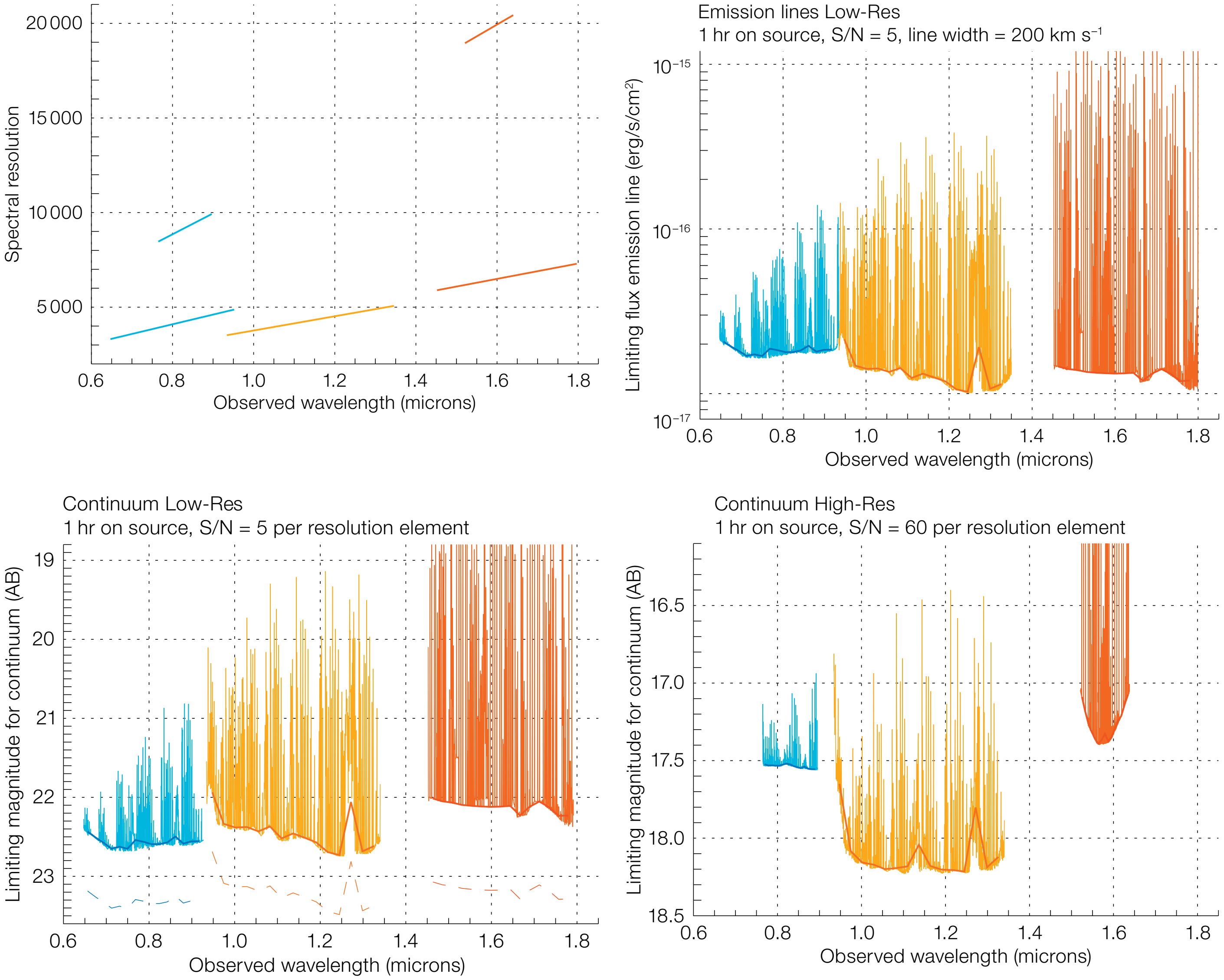

Figure 3. Top left: Spectral resolution $R = \lambda/\Delta\lambda$ as a function of wavelength for the low-resolution and high-resolution modes. In all the panels showing sensitivity, the thick solid lines show the typical value outside strong OH sky lines. At the resolution of MOONS more than 60–70% of the observed regions are completely free from OH sky lines. Top right: limiting flux for emission lines in low-resolution mode in the three simultaneous channels, for 1 hour on-source integration with S/N = 5 at the line peak. Bottom left: limiting magnitude in continuum for low-resolution mode in the three simultaneous channels, for 1 hour on-source integration with S/N = 5 per resolution element (~ 3 pixels) and dashed lines when rebinned to a resolution R = 1000 after sky subtraction. Bottom right: limiting magnitude in continuum for high-resolution mode in the three simultaneous channels, for 1 hour on source integration with S/N = 60 per resolution element (~ 3 pixels).

## Observing strategies and sky subtraction

Accurate subtraction of the sky background is critical when observing faint sources, particularly in the near-infrared, where strong OH sky lines dominate the background. To achieve this goal, we have implemented multiple methods. First of all, the spectral resolving power of $R > 4000$ for the medium-resolution mode ($R > 6500$ in the *H* band) ensures that at least 60–70% of the observed regions in the *Y*, *J* or *H* bands are completely free from OH airglow. Sky subtraction with fibres is challenging since their efficiency might change (even slightly) when they move. For this reason, particular attention has been devoted during the manufacturing of the fibres and their routing within the instrument to minimise the variation of focal ratio degradation (FRD), which has been measured to be << 1%. In order to remove any residuals, it is also possible to obtain a fast attached flat after the fibres have been reconfigured and are in their science position. To further optimise the sky subtraction, the fibre positioners have been designed to have overlapping patrol fields and are capable of being placed at a distance of 10 arcseconds from each other, so as to sample the sky very close to the science target.

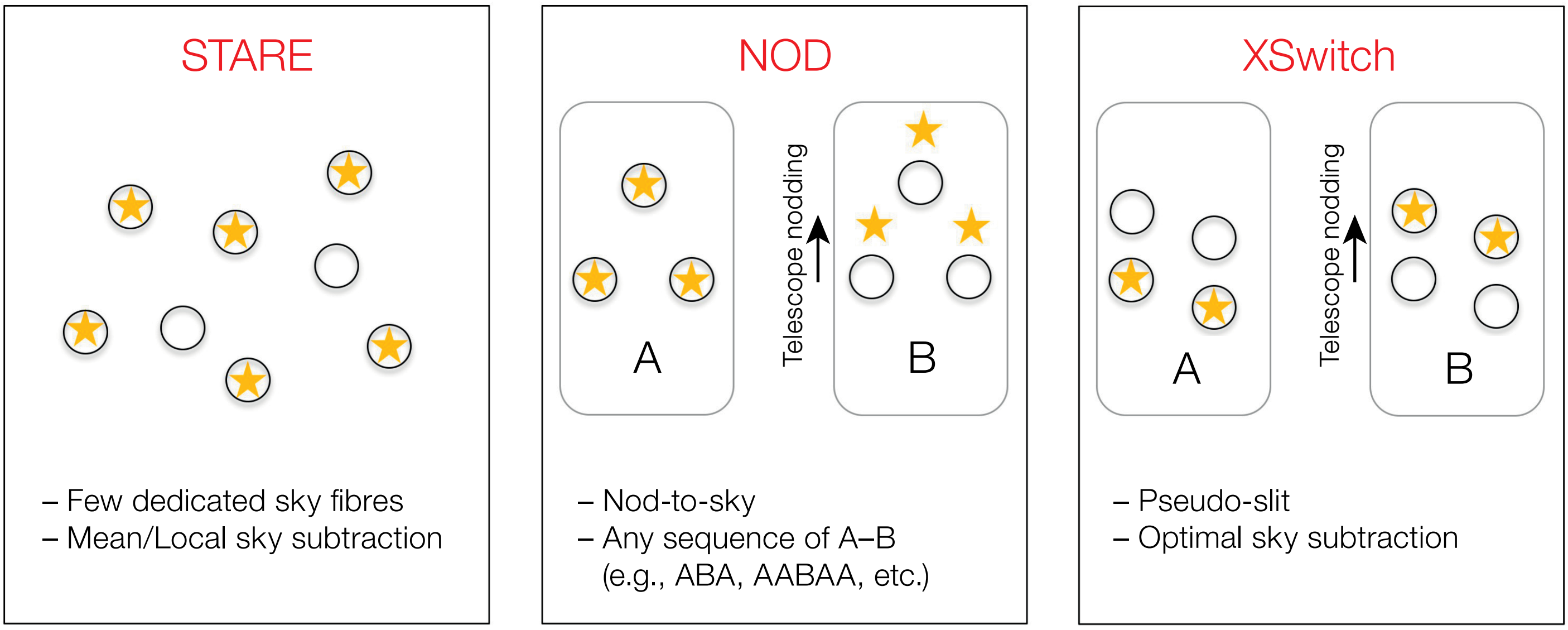

Figure 4. The three possible MOONS observing strategies envisaged for sky subtraction.

These considerations are driving the three possible observing strategies envisaged for MOONS (see list below and Figure 4). During commissioning these strategies will be tested and the performance of sky subtraction evaluated in order to provide guidelines to users.

1. *Stare:* The vast majority of fibres will be on the targets, with dedicated sky fibres distributed across the focal plane. The number of sky fibres can be determined by the user.
2. *Stare+nod:* The majority of, if not all, fibres will be on the targets and the telescope is then nodded to a nearby sky position. This has the advantage that the sky flux will pass through the same fibre as the target, thus removing many instrumental effects. The quality of sky subtraction will depend on the frequency of sky nods.
3. *XSwitch:* This provides a pseudo-slit observation with the most accurate sky subtraction. Every science fibre will have an adjacent sky fibre at the same fixed distance (10 < *d* < 30 arcseconds) and same direction. The telescope is then nodded by the same distance and direction, so that object and sky fibres are reversed. This observing pattern strategy allows both temporal and spatial sky variation to be removed, as well as accounting for instrumental effects (see Rodrigues et al., 2012 for more details and on-sky testing of this strategy).

## The Consortium

Reflecting the wide range of science goals, the MOONS Consortium builds on the scientific and technical expertise of a range of institutes in Chile, France, Germany, Italy, Norway, Portugal, Switzerland, the United Kingdom, and ESO. It includes ~ 100 engineers and 150 scientists across ~ 50 institutes. Table 2 shows the main roles of each of the institutes involved in the construction of the instrument.

## Schedule

The MOONS project passed the Final Design Review (FDR) in 2017 and is now fully in the assembly integration and verification (AIV) phase. The vast majority of the components have been manufactured and are now being integrated in Edinburgh. The Provisional Acceptance in Europe (PAE) is foreseen for the end of 2021, followed by the installation and commissioning at the VLT at the beginning of 2022 (see Figure 5).

## Links

[1] The official MOONS website: www.vltmoons.org
[2] The MOONS website at ESO: https://www.eso.org/sci/facilities/develop/instruments/MOONS.html
[3] See the fibre positioning units in action at https://vltmoons.org/resources/.

## Notes

[a] For information contact Michele Cirasuolo at mciras@eso.org.
[b] A short anecdote: A very small fraction of the light coming from the fibres does not reach the detector immediately but bounces back and forth between the optical surfaces and when it reaches the detector it creates a “ghost” image! That image is adding noise and therefore degrading the science performance. When we discovered this problem, we had to think how to remove this effect. In order to do this each slitlet was equipped with a special component, tilted with respect to the optical axis, to deviate the “ghost” away from the detector. During a telecon someone said: “it works like an exorcist; the ghost has been evicted”. Since then this special component is called “the Exorcist”!

Table 2. Role of the consortium construction partners.

| Institute | Work package | Logos |
|---|---|---|
| STFC UK Astronomy Technology Centre Edinburgh | Project office, fibre positioning units, calibration unit, cryostat, detector adjustment module, AIV, control software | |
| Cambridge University | Camera opto-mechanics, assembly and testing | |
| Eidgenössische Technische Hochschule (ETH) Zürich | Fibre positioning unit | |
| INAF – Firenze | Optical design, exchange VPH mechanisms | |
| INAF – Roma | Acquisition cameras end-to-end modelling | |
| INAF – Milano | Observation preparation software and path analysis | |
| GEPI – Paris | Fibre assembly, slit and shutters, data reduction software | |
| University of Geneva | Instrument control electronics | |
| Instituto de Astrofísica e Ciências do Espaço | Field corrector, rotating front end structure, cable wrap | |
| Pontificia Universidad Católica de Chile | Metrology system, instrument control software | |
| ESO | Detector arrays and CCDs | |

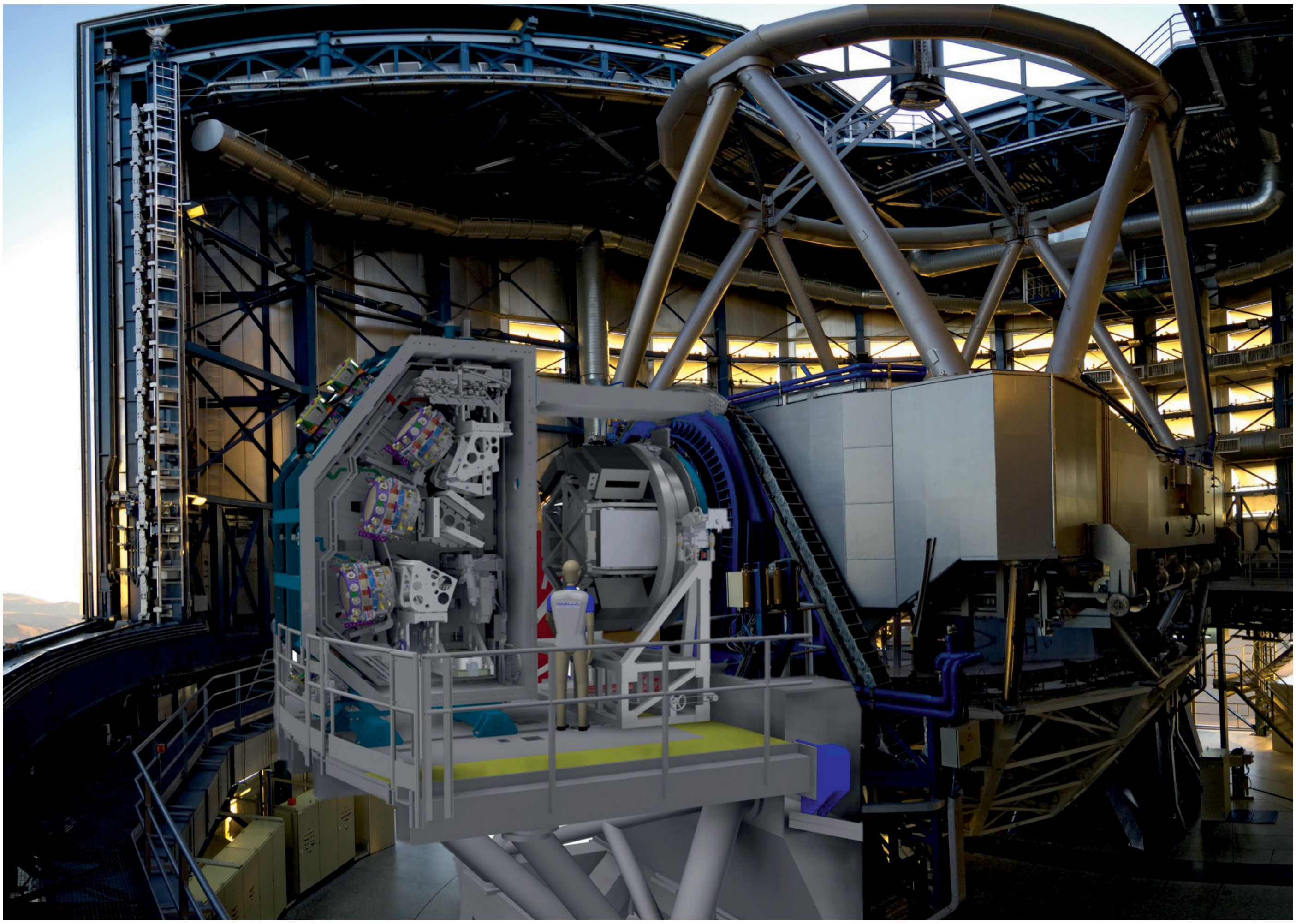

Figure 5. Artist's impression of MOONS on the Nasmyth platform at the VLT.